\documentclass[aps,prx,twocolumn]{revtex4}

\usepackage{dcolumn}
\usepackage{bm}
\usepackage{amsmath}
\usepackage{graphicx}
\usepackage{amsmath,amssymb}

\newcommand{\pro}[2]{\langle{#1}|{#2}\rangle}
\newcommand{\bra}[1]{\langle{#1}|}
\newcommand{\ket}[1]{|{#1}\rangle}

\begin{document}


\title{Quantum self-learning Monte Carlo with quantum Fourier transform sampler}

\author{Katsuhiro Endo$^1$, Taichi Nakamura$^2$, Keisuke Fujii$^{3, 4}$, 
and Naoki Yamamoto$^5$}
\affiliation{
1 Department of Mechanical Engineering , Keio University, 
Hiyoshi 3-14-1, Kohoku, Yokohama, Japan \\
2 Department of System Design Engineering, Keio University, 
Hiyoshi 3-14-1, Kohoku, Yokohama, Japan \\
3 Graduate School of Engineering Science, Osaka University,
1-3 Machikaneyama, Toyonaka, Osaka 560-8531, Japan \\
4 Center for Quantum Information and Quantum Biology, Institute for Open and 
Transdisciplinary Research Initiatives, Osaka University, Japan \\
5 Department of Applied Physics and Physico-Informatics, 
and Quantum Computing Center, Keio University, 
Hiyoshi 3-14-1, Kohoku, Yokohama, Japan
}
\date{\today}


\begin{abstract} 
The self-learning Metropolis-Hastings algorithm is a powerful Monte Carlo method that, 
with the help of machine learning, adaptively generates an easy-to-sample probability 
distribution for approximating a given hard-to-sample distribution. 
This paper provides a new self-learning Monte Carlo method that utilizes a quantum 
computer to output a proposal distribution. 
In particular, we show a novel subclass of this general scheme based on the quantum 
Fourier transform circuit; 
this sampler is classically simulable while having a certain advantage over conventional 
methods. 
The performance of this ``quantum inspired" algorithm is demonstrated by some numerical 
simulations. 
\end{abstract}





\maketitle



\section{Introduction}
\label{intro}

Monte Carlo (MC) simulation is a powerful statistical method that is generically applicable 
to compute statistical quantities of a given system by sampling random variables from its 
underlying probability distribution. 
A particularly efficient method is the Metropolis-Hastings (MH) algorithm \cite{Hastings}; 
this realizes a fast sampling from a target distribution via an appropriate acceptance/rejection 
filtering of the samples generated from an alternative proposal distribution which is easier 
to sample compared to the target one. 
Therefore, the most important task in this algorithm is to specify an appropriate proposal 
distribution satisfying the following three conditions; 
(i) it must be easy to sample, 
(ii) the corresponding probability can be effectively computed for judging acceptance/rejection 
of the sample, and 
(iii) it is rich in representation, meaning that it may lie near the target distribution. 
This is a long-standing challenging problem, but the recent rapid progress of machine learning 
enables us to take a circumventing pathway for the issue, the self-learning MC 
\cite{Liu,Wang,Albergo,Huitao}, which introduces a parametrized proposal distribution and 
updates the parameters during the sampling so that it is going to mimic the target one. 
Despite of its potential power thanks to the aid of machine learning, this approach has been 
demonstrated only with a few physical models; 
for example in Ref.~\cite{Liu}, for a target hard-to-sample Ising distribution, a parametrized 
proposal Ising distribution was applied to demonstrate the effectiveness of self-learning 
MC approach. 
To expand the scope of self-learning MC, we need a systematic method to design a 
parametrized proposal distribution that is generically applicable to a wide class of target 
distribution.

Now we turn our attention to quantum regime, with the hope that the quantum computing 
might provide us an effective means to attack the above-mentioned problem. 
In fact the so-called quantum supremacy holds for sampling problems; that is, a 
(non-universal) quantum computer can generate a probability distribution which is hard to 
sample via any classical (i.e., non-quantum) computer. 
Especially, the Boson sampling \cite{Aaronson} and Instantaneous Quantum Polynomial time 
computations \cite{IQP1,IQP2} are well known, the former of which is now even within reach 
of experimental demonstration \cite{Spring 2013,Tillmann 2013}. 
Moreover, a recent trend is to extend this idea to quantum learning supremacy 
\cite{Born supremacy}, meaning that a quantum circuit is trained to learn a given target 
distribution faster than any classical computer.

With the above background in mind, in this paper we study a new type of self-learning 
MC that uses quantum computing to generate a proposal distribution. 
In fact this scheme satisfies the above-described three conditions. 
First, (i) is already fulfilled as an intrinsic nature of quantum computers. 
Second, it is well known that the task (ii) can be effectively executed using the amplitude 
estimation algorithm \cite{Brassard,Suzuki}, which is in fact twice as fast as the classical 
correspondence. 
Lastly the above-described fact, the expressive power of quantum computers for generating 
complex probability distributions, might enable us to satisfy (iii) and mimic a target distribution 
which is essentially hard to sample via any classical means. 
To realize the learning scheme in a quantum system, we take the variational method, 
meaning that a parametrized quantum circuit is trained so that its output probability 
distribution approaches to the target distribution. 
This schematic itself is also employed in the quantum generative modeling 
\cite{Generative model 1,Generative model 2}, but application to MH might be of more 
practical use for the following reason. 
That is, unlike the generative modeling problem, MH need not generate a proposal 
distribution that is very close to the target, but rather it requires only a relatively high 
acceptance ratio and accordingly less demanding quantum computers.

Of course the most difficult part is to design a parametrized quantum circuit which may 
successfully generate a suitable proposal distribution. 
Therefore, for the purpose of demonstrating the proof-of-concept, in this work we consider 
a special type of quantum circuit composed of the quantum Fourier transform (QFT), where 
the parameters to be learned are assigned corresponding to only the low-frequencies 
components. 
In fact, thanks to the rich expressive power of the Fourier transform in representing or 
approximating various functions, the proposed QFT sampler is expected to satisfy the 
condition (iii), in addition to (i) and (ii). 
We also emphasize that this QFT sampler or its variant (e.g., with different parametrization 
to cover the high-frequencies components) is a new type of circuit ansatz in the quantum 
variational method and might be applicable to other problems such as the quantum generative 
modeling.

Now we state our bonus theorem; the proposed QFT sampler can be efficiently simulated 
with a classical means, using the iterative measurement technique \cite{Browne}. 
This is exactly the direction to explore a classical algorithm that fully makes use of quantum 
feature, i.e., a quantum-inspired algorithm such as \cite{Tang}. 
Actually we will show that this quantum-inspired sampler has a certain advantage over 
some conventional methods, in addition to the clear merit that the system with e.g., 
hundreds of qubits is simulable.

We use the following notations: 
for a complex column vector $\bm a$, the symbols $\bm a^\dagger$ and $\bm a^\top$ 
represent its complex conjugate transpose and transpose vectors, respectively. 
Also $\bm a^*$ denotes the elementwise complex column vector of $\bm a$. 
Hence $\bm a^\dagger = (\bm a^*)^\top$.


\section{General scheme of quantum self-learning Monte Carlo method}


\subsection{Classical MH algorithm}

Let $p({\bm x})$ and $q({\bm x})$ be target and proposal probability distributions, respectively. 
To get a sample from $p({\bm x})$, the MH algorithm instead samples from $q({\bm x})$ and 
accepts the result with valid probability determined by the detailed balance conditions. 
More specifically, assume that we have last accepted a sample ${\bm r}$ and now obtain 
a sample $\tilde{\bm r}$ generated from the proposal distribution $q({\bm x})$. 
Then, this sample $\tilde{\bm r}$ is accepted with probability 
\begin{equation}
\label{acceptance ratio}
      A({\bm r}, \tilde{\bm r}) 
           = {\rm min}\left\{ 1,  ~ \frac{p(\tilde{\bm r})q({\bm r})}{p({\bm r})q(\tilde{\bm r})} \right\},
\end{equation}
which is called the acceptance ratio. 
Note that the value $p(\bm r)$ is assumed to be easily computable for a given $\bm r$, 
while its sampling is hard. 
This procedure is repeated until the number of samples becomes enough large; 
then these accepted samples are governed by the target distribution $p({\bm x})$ due to 
the detailed balance conditions. 
Note that, if $q({\bm x})=p({\bm x})$, then the acceptance ratio is always exactly $1$, which 
is maximally efficient; 
but of course this does not happen because $p({\bm x})$ is hard to sample while $q({\bm x})$ 
is assumed to be relatively easy to sample.

In the context of self-learning MC, a parametric model of the proposal distribution 
$q({\bm x};{\bm\theta})$ is considered, with ${\bm\theta}$ the vector of parameters. 
The self-learning MC aims to learn the parameters so that $q({\bm x};{\bm\theta})$ well 
approximates the target $p({\bm x})$.


\subsection{General form of the quantum self-learning MH algorithm}

Here we describe the quantum sampler executing the self-learning MH algorithm, in the 
general setting. 
First, for the initial state $\ket{g^N}=\ket{g}^{\otimes N}$ with $\ket{g}=[1,0]^\top$ 
a qubit state, we apply the parametric unitary gate $U({\bm\theta})$:
\[
      \ket{\Psi(\bm\theta)} = U({\bm\theta}) \ket{g^N}, 
\]
where $\bm \theta\in{\mathbb C}^m$ are the parameters to be tuned. 
The reason of taking the complex-valued parameters will be made clear in the next section 
when specializing to the QFT circuit. 
This state is measured in the computational basis, defining the probability distribution 
\begin{equation}
\label{proposal dis}
     q(\bm x;{\bm\theta}) 
          = | \pro{\bm x}{\Psi(\bm\theta)} |^2
          = | \bra{\bm x}U({\bm\theta}) \ket{g^N} |^2, 
\end{equation}
where $\bm x$ is the multi-dimensional random variable represented by binaries 
(an example is given in Appendix~A). 
Equation \eqref{proposal dis} is the proposal distribution of our MH algorithm. 
Hence, our goal is to update $\bm\theta$ so that $q(\bm x;{\bm\theta})$ is going to well 
approximate the target distribution $p({\bm x})$. 
The learning process for updating $\bm\theta$ is executed via the standard gradient 
descent method of a loss function, as described below.

First, for a fixed $\bm \theta$, we obtain samples ${\bm r_1, \ldots, \bm r_B}$ from 
$q(\bm x;{\bm\theta})$ by the computational-basis measurement. 
These samples are filtered according to the acceptance probability \eqref{acceptance ratio}, 
thus producing samples governed by $p(\bm x)$. 
Note now that, for a given $\bm r$, the value $q(\bm r) = | \pro{\bm r}{\Psi(\bm\theta)} |^2$ 
must be effectively computed to do this filtering process (recall that the value of $p(\bm r)$ 
is assumed to be easily obtained); 
this computability indeed depends on the structure of $U({\bm\theta})$, but in general we 
could apply the quantum amplitude estimation algorithm \cite{Brassard,Suzuki} to reduce 
the cost for executing this task. 
The samples ${\bm r_1, \ldots, \bm r_B}$ are also used to calculate the gradient descent 
vector of the loss function $L(\bm\theta)$ for updating $\bm\theta$. 
Noting that $L(\bm\theta)$ is real while $\bm\theta$ is complex, its infinitesimal change 
with respect to $\bm\theta$ is given by 
\[
      \delta L = \Big( \frac{\partial L}{\partial \bm\theta} \Big)^\top \delta\bm\theta
                   + \Big( \frac{\partial L}{\partial \bm\theta^*} \Big)^\top \delta\bm\theta^*
                  = \Big( \frac{\partial L}{\partial \bm\theta} \Big)^\top \delta\bm\theta
                   + \Big( \frac{\partial L}{\partial \bm\theta} \Big)^\dagger \delta\bm\theta^*. 
\]
The gradient descent vector for updating the parameter from $\bm \theta$ to 
$\bm\theta' = \bm\theta + \delta\bm\theta$ is thus given by 
\begin{equation}
\label{standard grad}
     \bm\theta' = \bm\theta 
              - \alpha \Big(\frac{\partial L(\bm\theta)}{\partial \bm\theta}\Big)^*, 
\end{equation}
where $\alpha>0$ is the learning coefficient; 
in fact then $\delta L=-2\alpha \| \partial L/\partial \bm\theta \|^2 \leq 0$. 
In this work, the loss function is set to the following cross entropy between 
$q(\bm x;\bm\theta)$ and $p(\bm x)$: 
\begin{equation}
\label{cross entropy}
     L(\bm\theta) = - \sum_{\bm x} p(\bm x) \log{ q(\bm x;\bm\theta) },
\end{equation}
which is a standard measure for quantify the similarity of two distributions. 
The gradient vector of $L(\bm\theta)$ can be computed as follows:
\begin{eqnarray}
\label{general gradient}
& & \hspace*{-2.1em}
        \frac{\partial L(\bm\theta)}{\partial \bm\theta} 
                =  - \sum_{\bm x} 
                     \frac{p(\bm x)}{q(\bm x;\bm\theta)}
                     \frac{\partial q(\bm x; \bm\theta)}{\partial \bm\theta} 
\nonumber \\ & & \hspace*{1em}
                =  - \sum_{\bm x} q(\bm x;\bm\theta)
                     \frac{p(\bm x)}{q(\bm x;\bm\theta)^2}
                     \frac{\partial q(\bm x; \bm\theta)}{\partial \bm\theta} 
\nonumber \\ & & \hspace*{1em}
                =  \lim_{n\to\infty} - \frac{1}{n}\sum_{i=1}^n 
                     \frac{p(\bm r_i)}{q(\bm r_i;\bm\theta)^2}
                     \frac{\partial q(\bm r_i; \bm\theta)}{\partial \bm\theta} 
\nonumber \\ & & \hspace*{1em}
        \simeq  - \frac{1}{B} \sum_{i=1}^B
                     \frac{p(\bm r_i)}{q(\bm r_i;\bm\theta)^2}
                     \frac{\partial q(\bm r_i; \bm\theta)}{\partial \bm\theta}. 
\end{eqnarray}
Note that 
\[
      \frac{\partial q(\bm r; \bm\theta)}{\partial \bm\theta}
      = \left[ |\bra{\bm r} \frac{\partial U(\bm\theta)}{\partial \theta_1} \ket{g^N} |^2,
            \ldots,
            |\bra{\bm r} \frac{\partial U(\bm\theta)}{\partial \theta_m} \ket{g^N} |^2  \right]
\]
can be directly estimated when $U(\bm\theta)$ is composed of Pauli operators with 
the parameters $\{\theta_i\}$ corresponding to the rotation angles \cite{Mitarai 2018}.

Here we discuss the notable feature and possible quantum advantage of the quantum 
sampler for the MH algorithm, by referring to the three conditions mentioned in 
Sec.~\ref{intro}. 
First, the condition (i) is indeed satisfied because now the sampler is a quantum device 
that physically produces each measurement result ${\bm r}$ according to the proposal 
probability distribution $q({\bm x};{\bm\theta})$, only in a few micro second in the case of 
superconducting devices. 
As for the condition (ii), it is in principle possible to effectively compute the probability 
$q({\bm r};{\bm\theta})$ for a given ${\bm r}$, as discussed above in this subsection. 
Lastly for the condition (iii), $q({\bm x};{\bm\theta})$ might be able to represent a wide 
class of probability distribution, which is even hard to sample via any classical means as 
mentioned in Sec.~\ref{intro}. 
Realization of this possible quantum advantage of course needs a clever designing of 
the ansatz $U(\bm\theta)$.


\section{The quantum Fourier transform sampler}

To show the proof of principle of the quantum sampler for self-learning MH algorithm, 
here we consider a special class of circuit composed of QFT, called the QFT sampler. 
Importantly, as will be shown, the QFT sampler is classically simulable, while it might have 
an advantage over classical algorithms.


\subsection{1-dimensional QFT sampler}

\begin{figure}[htp]
    \centering
    \includegraphics[width=0.8\linewidth]{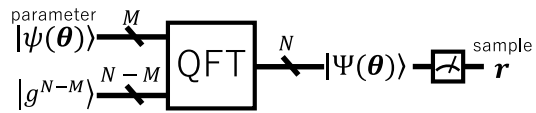}
    \caption{ Schematic illustration of the quantum Fourier transform sampler． }
    \label{fig:m1}
\end{figure}

We begin with the 1-dimensional QFT sampler; extension to the multi-dimensional case 
is discussed in the next subsection. 
As illustrated in Fig.\ref{fig:m1}, this sampler is composed of the QFT operation applied to 
a $N$-qubits input state $\ket{\rm in}= \ket{{\psi}({\bm\theta})} \otimes \ket{g^{N-M}}$ where
\begin{equation}
\label{QFT input state}
       \ket{{\psi}({\bm\theta})}
          =\theta_0 \ket{0} + \theta_1 \ket{1} + \cdots + \theta_{2^M-1} \ket{2^M-1}, 
\end{equation}
where $\{\ket{0}, \ket{1}, \ldots, \ket{2^M-1}\}$ is the set of computational basis states in 
$({\mathbb C}^2)^{\otimes M}={\mathbb C}^{2^M}$, e.g., $\ket{0}=\ket{g^M}$. 
That is, the first $M$-qubits state contains the parameters 
${\bm\theta}=[\theta_0, \cdots, \theta_{2^M-1}]^\top \in {\mathbb C}^{2^M}$, while the 
residual $N-M$ qubits are set to $\ket{g}$ states. 
Note that, if ${\bm\theta} \in {\mathbb R}^{2^M}$, then the proposal distribution 
\eqref{proposal dis QFT} below is limited to an even function, and thus ${\bm\theta}$ 
must take complex numbers. 
The output of QFT is given by $\ket{\Psi(\bm \theta)}=U_{\rm QFT}\ket{{\rm in}}$, 
where $U_{\rm QFT}$ is the QFT unitary operator whose matrix representation is given by 
\[
        \bra{k}U_{\rm QFT}\ket{j} = \frac{1}{\sqrt{2^N}}e^{i2\pi kj/2^N}. 
\]
Now the measurement on $\ket{\Psi(\bm \theta)}$ in the computational basis yields 
the probability distribution 
\begin{equation}
\label{proposal dis QFT}
     q_{\rm QFT}(x;{\bm\theta}) = |\pro{x}{\Psi(\bm \theta)}|^2 
                = | \bra{x}U_{\rm QFT}\ket{{\rm in}} |^2, 
\end{equation}
where the random variable $x$ is represented with binaries, i.e., $x \in \{0,1,\cdots, 2^N-1\}$.

Again, the task of self-learning MH is to update $\bm \theta$ so that the proposal 
distribution $q_{\rm QFT}(x;{\bm\theta})$ may approach to the target distribution $p(x)$. 
To compute the gradient vector \eqref{general gradient}, let us express 
$\bra{x}U_{\rm QFT}\ket{\rm in}$ as 
\begin{equation}
\label{inner product}
       \bra{x}U_{\rm QFT}\ket{\rm in} = {\bm u}_x^\top {\bm\theta}, 
\end{equation}
where ${\bm u}_x$ is the vector composed of the first $2^M$ elements of the $x$-th row 
vector of $U_{\rm QFT}$; hence the $j$th element of ${\bm u}_x$ is given by 
\[
       ({\bm u}_{x})_j = \frac{1}{\sqrt{2^N}}e^{i2\pi xj/2^N} (j=0, 1, \cdots, 2^M-1).
\]
Then the partial derivative of 
$q_{\rm QFT}(x;\bm\theta) = ({\bm u}_x^\top {\bm\theta})^*({\bm u}_x^\top {\bm\theta})
= ({\bm u}_x^\dagger{\bm\theta}^*)({\bm u}_x^\top {\bm\theta})$ with respect to 
$\bm\theta$ is given by 
\[
       \frac{\partial q_{\rm QFT}(x;\bm\theta)}{\partial \bm\theta} 
            = ({\bm u}_x^\dagger{\bm\theta}^*) {\bm u}_x. 
\]
Hence, from Eq.~\eqref{general gradient}, the gradient vector of the loss function 
$L(\bm\theta)$ can be calculated as follows:
\begin{eqnarray}
& & \hspace*{-2.1em}
        \Big( \frac{\partial L(\bm\theta)}{\partial \bm\theta} \Big)^*
          \simeq  -\frac{1}{B} \sum_{i=1}^B 
                       \frac{p(r_i)}{q_{\rm QFT}(r_i;\bm\theta)^2} 
                           ({\bm u}_{r_i}^\top {\bm\theta}){\bm u}_{r_i}^*,
\label{compute gradient}
\end{eqnarray}
where $\{r_1,r_2, \cdots, r_B\}$ are samples taken from the proposal distribution 
$q_{\rm QFT}(x;\bm\theta)$. 
Recall that we filter these samples via the rule \eqref{acceptance ratio} and thereby obtain 
samples that are subjected to the target distribution $p(x)$. 
In this work, instead of the standard gradient update \eqref{standard grad}, we take the 
following momentum gradient descent \cite{Momentum grad}; 
\begin{equation}
\label{momentum method}
     \bm\theta' = \bm\theta - \alpha \bm m', ~~~
     \bm m' = \mu \bm m + (1-\mu) 
                                     \Big(\frac{\partial L(\bm\theta)}{\partial \bm\theta}\Big)^*, 
\end{equation}
where $\alpha$ and $\mu$ are the learning coefficients. 
The updated vector $\bm\theta'$ is normalized and substituted into Eq.~\eqref{QFT input state} 
for the next learning stage. 
Note that Eq.~\eqref{momentum method} is identical to the standard gradient descent 
when $\mu=0$.

Here let us discuss how the basic conditions (i)-(iii) are reasonably fulfilled by the QFT sampler. 
First, as mentioned before, the QFT sampler enables us to obtain samples enough fast, thanks 
to the intrinsic feature of quantum devices, and thus it satisfies the condition (i). 
Also the probability $q_{\rm QFT}(r;{\bm\theta})$ with given $r$ is obtained via just calculating 
the inner product \eqref{inner product} which costs of the order $O(2^M)$, thereby the 
condition (ii) is satisfied when $2^M$ is a classically tractable number. 
As for the condition (iii), noting that the parameterized ket vector $\ket{\psi(\bm\theta)}$ 
serves as the low-frequency components of the shape of the proposal distribution 
$q_{\rm QFT}(x;{\bm\theta})$, the QFT sampler may be able to well approximate the shape 
of the target $p(x)$ in view of the rich expressive power of Fourier decomposition.

Now, one might think that this QFT sampler is still out of reach, even in the case of medium-size 
quantum devices with e.g., $N=100$ qubits. 
However, remarkably, the QFT sampler can be realized in a classical digital computer as long as 
$N^2$ and $2^M$ are classically tractable numbers; 
that is, in this regime, this is a ``quantum-inspired" algorithm that can deal with even a random 
variable on $2^N=2^{100}$ discrete elements. 
The trick relies on the use of the adaptive measurement technique \cite{Browne}, which 
enables us to sample only by applying $O(2^M + N)$ operations in a classical computer; 
see Appendix~B for a detailed explanation. 
This means that, therefore, the QFT sampler fulfills the condition (i), even as a classical 
computer.

Finally, we discuss a possible advantage of our QFT sampler, within a regime of classical 
sampling method. 
As a classical Fourier-based proposal distribution, one might think to employ the fast 
Fourier transform (FFT). 
However, to deal with a variable with $2^N$ discretized elements, FFT needs ${\cal O}(N2^N)$ 
operations, while QFT can realize the same operation only with ${\cal O}(N^2)$ gates. 
As is well known, this does not mean a quantum advantage in the typical application scene 
such as signal processing, because all the amplitude of the QFT-transformed state cannot 
be effectively determined \cite{Nielsen and Chuang}. 
On the other hand, the presented scheme only requires sampling and thus determining 
$\{ q_{\rm QFT}(r_i;{\bm\theta}) \}_{i=1,\ldots, B}$ rather than all the elements 
$\{ q_{\rm QFT}(x;{\bm\theta}) \}_{x=0,\ldots, 2^N-1}$. 
Hence we could say that the developed quantum-inspired algorithm has a solid computational 
advantage over the known classical algorithm, in the problem of determining a Fourier-based 
proposal distribution for the MH algorithm.


\subsection{Multistage QFT sampler for multi-dimensional distributions}
\label{sec:multi QFT sampler}

\begin{figure}[htp]
    \centering
    \includegraphics[width=1.0\linewidth]{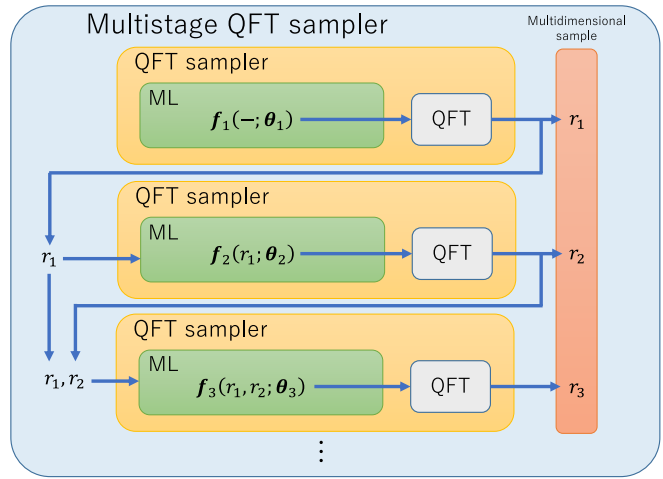}
    \caption{ Schematic illustration of the multistage QFT sampler, where ML means 
    machine learning. }
    \label{fig:m2}
\end{figure}

The QFT sampler discussed above is able to sample only from a 1-dimensional distribution. 
To extend the scheme to the case of multi-dimensional distributions, we here develop a 
multistage QFT sampler, which is composed of single QFT samplers with their parameters 
updated via machine learning, as illustrated in Fig.~\ref{fig:m2}.

First, to see the idea, let us consider the case where the $D$ random variables $x_1, \cdots, x_D$ 
are independent with each other and subjected to the $D$-dimensional independent 
target distribution $p({\bm x})$. 
In this case, the following proposal distribution might work: 
\begin{equation}
\label{independent target}
     q({\bm x} ; \bm\theta) = \prod_{k=1}^{D} q_{\rm QFT}(x_k ; \bm\theta_k),
\end{equation}
where $\bm x=[x_1, \cdots, x_D]^\top$ is the $D$-dimensional random variable 
represented by binaries $x_k \in \{0, 1,\cdots, 2^N-1\}$. 
Here $q_{\rm QFT}(x_k ; \bm\theta_k)$ is a 1-dimensional QFT sampler parametrized by the 
$2^M$-dimensional complex vector $\bm\theta_k$; 
we summarize these vectors to ${\bm\theta}=[\bm\theta_1^\top, \cdots, \bm\theta_D^\top ]^\top$.

Now based on Eq.~\eqref{independent target} we construct the multistage QFT sampler 
that can deal with a non-independent multi-dimensional proposal distribution. 
The point is that, as illustrated in Fig.~\ref{fig:m2}, the parameter vector $\bm\theta_k$ 
specifying the $k$-th 1-dimensional QFT sampler in Eq.~\eqref{independent target} is 
replaced by a vector of parametrized functions of random variables up to the $(k-1)$-th 
stage, i.e., ${\bm f}_k( x_1, \cdots, x_{k-1} ; \bm\theta_k )$, whose control parameter 
$\bm\theta_k$ is to be repeatedly modified through the learning process. 
Hence the proposal distribution is given by 
\begin{equation}
\label{D dim proposal}
        q({\bm x};\bm\theta) 
         = \prod_{k=1}^{D}  q_{\rm QFT}\left( x_k ; {\bm f}_k( x_1, \cdots, x_{k-1} ; \bm\theta_k ) \right).
\end{equation}
Note that this is simply a representation of the joint probability distribution over the 
multi-dimensional random variables $\bm x=[x_1, \cdots, x_D]^\top$ via the series of 
conditional probabilities. 
The gradient of the cross entropy \eqref{cross entropy} is derived in the same way as 
the 1-dimensional case; 
\begin{equation}
\label{multi dim gradient}
     \Big(\frac{\partial L(\bm\theta)}{\partial \bm\theta_k}\Big)^* 
         \simeq
              -\frac{1}{B} \sum_{i=1}^B \frac{p(\bm r_i)}{q({\bm r}_i;\bm\theta)}
                \frac{({\bm u}_{r_k}^\top {{\bm f}_k}){\bm u}_{r_k}^*}{ q_{\rm QFT}(r_k ; {\bm f}_k) }
                \frac{ \partial {\bm f}_k }{ \partial \bm\theta_k }, 
\end{equation}
where $\{ \bm r_1, \cdots, \bm r_B \}$ are samples produced from the proposal distribution 
$q({\bm x};\bm\theta)$. 
Note that each sample $\bm r=[r_1, \cdots, r_D]^\top$ is formed from $r_k$ produced from 
the $k$-th 1-dimensional QFT sampler, as shown in Fig.~\ref{fig:m2}. 
Also, as in the 1-dimensional case \eqref{inner product}, the vector ${\bm u}_{x_k}$ is defined 
through $\bra{x_k}U_{\rm QFT}\ket{\rm in} 
= {\bm u}_{x_k}^\top {\bm f}_k( x_1, \cdots, x_{k-1} ; \bm\theta_k )$. 
This gradient vector \eqref{multi dim gradient} is used to update each parameter vector 
$\bm \theta_k$ using the momentum gradient descent \eqref{momentum method}, which 
is now of the form 
\[
     \bm\theta_k' = \bm\theta_k - \alpha \bm m_k', ~~~
     \bm m_k' = \mu \bm m_k + (1-\mu) 
                                     \Big(\frac{\partial L(\bm\theta)}{\partial \bm\theta_k}\Big)^*, 
\]
where the learning coefficients $(\alpha, \mu)$ do not depend on $k$ for simplicity. 
This eventually constitutes the total gradient descent vector minimizing the loss function 
\eqref{cross entropy} and accordingly move the proposal distribution \eqref{D dim proposal} 
toward the target $p(\bm x)$. 
Also recall that we use Eq.~\eqref{acceptance ratio} to filter $\{ \bm r_1, \cdots, \bm r_B \}$ 
to obtain samples that are subjected to $p(\bm x)$.

In this work we examine the following four models of the parameterized function 
${\bm f}_k( x_1, \cdots, x_{k-1} ; \bm\theta_k )$. 
\begin{itemize}

\item Identity (Id) model: 
\[
     {\bm f}_k( x_1, \cdots, x_{k-1} ; \bm\theta_k ) 
       = {\rm Norm}\left( \bm\theta_k \right) 
       = \frac{\bm\theta_k}{ \| \bm\theta_k \| }, 
\]
where $\| \bm \theta \|=\sqrt{ \bm\theta^\dagger \bm\theta}$. 
The ${\rm Norm}$ function ensures the normalization of $\ket{{\bm \psi}({\bm\theta})}$. 
This leads to the independent proposal distribution \eqref{independent target}.

\item Linear Basis Linear Regression (LBLR) model:
\begin{eqnarray}
& & \hspace*{-0.6em}
     {\bm f}_k( x_1, \cdots, x_{k-1} ; \bm\theta_k ) 
\nonumber \\ & & \hspace*{-0.5em}
     = {\rm Norm}\left(  {\bm w}_1^{(k)} x_1 + \cdots 
              + {\bm w}_{k-1}^{(k)} x_{k-1} + \bm b^{(k)} \right),
\nonumber
\end{eqnarray}
where in this case the parameters to be learned are the collection of $2^M$-dimensional 
complex vectors, 
$\bm\theta_k = \{\bm w_1^{(k)}, \bm w_2^{(k)}, \cdots, \bm w_{k-1}^{(k)}, \bm b^{(k)} \}$. 
This model outputs a linear combination of their input argument.

\item Nonlinear Basis Linear Regression (NBLR) Model: 
\begin{eqnarray}
& & \hspace*{-0.6em}
     {\bm f}_k( x_1, \cdots, x_{k-1} ; \bm\theta_k ) 
\nonumber \\ & & \hspace*{0em}
     = {\rm Norm}\left(  \sum_{j=1}^J  {\bm w}_{j}^{(k)} \phi_j^{(k)}(x_1, \cdots, x_{k-1}) 
                                + \bm b^{(k)} \right),
\nonumber
\end{eqnarray}
where $\bm\theta_k = \{\bm w_1^{(k)},\bm w_2^{(k)}, \cdots, \bm w_J^{(k)}, \bm b^{(k)} \}$ are 
the collection of $2^M$-dimensional complex vectors to be learned, and $\{\phi_j^{(k)}(\cdot)\}$ 
are fixed nonlinear basis functions. 
Different set of nonlinear basis functions should be chosen for a different target distribution.

\item Neural Network (NN) model:
\begin{eqnarray}
& & \hspace*{-0.6em}
     {\bm f}_k( x_1, \cdots, x_{k-1} ; \bm\theta_k ) 
        = {\rm Norm}\big(  h(\bm y^{(k)}) \big), 
\nonumber \\ & & \hspace*{-0.5em}
      \bm y^{(k)} = g_4^{(k)} \circ {\rm Sg} \circ g_3^{(k)} \circ 
                 {\rm Sg} \circ g_2^{(k)} \circ {\rm Sg} \circ g_1^{(k)} (\bm x), 
\nonumber
\end{eqnarray}
where $\circ$ denotes $g_1\circ g_2(\bm x)=g_1(g_2(\bm x))$. 
This is a fully connected 4-layers neural network with input $\bm x=(x_1, \cdots, x_{k-1})$, 
shown in Fig.~\ref{fig:densenet}. 
Here $g_j^{(k)}(\bm x) = {\bm W}_j^{(k)} \bm x + \bm b_j^{(k)}$ is a linear transformation 
between the layers; 
$({\bm W}_2^{(k)}, {\bm W}_3^{(k)})$ are $256\times 256$ real parameter matrices, and 
$({\bm W}_1^{(k)}, {\bm W}_4^{(k)})$ are $256\times (k-1)$ and $2^{M+1} \times 256$ real 
parameter matrices, respectively; 
also $({\bm b}_1^{(k)}, {\bm b}_2^{(k)}, {\bm b}_3^{(k)})$ are $256$ dimensional real parameter 
vectors, and ${\bm b}_4^{(k)}$ is a $2^{M+1}$ dimensional real parameter vector. 
Hence 
$\bm\theta_k = (\bm W_1^{(k)}, \cdots, \bm W_4^{(k)}, \bm b_1^{(k)}, \cdots, \bm b_4^{(k)})$ 
are the parameters to be optimized. 
${\rm Sg}(\bm x)$ is the sigmoid function whose $\ell$-th component is defined as 
$1/(1+e^{-x_\ell})$. 
Finally, for the output vector $\bm y=[\bm y_1^\top, \bm y_2^\top]^\top$ where $\bm y_1$ 
and $\bm y_2$ are the $2^M$ real vectors, the function $h$ acts on it and produces 
a complex vector $h(\bm y)=\bm y_1 + i\bm y_2$. 

\begin{figure}[h]
    \centering
    \includegraphics[height=0.45\linewidth]{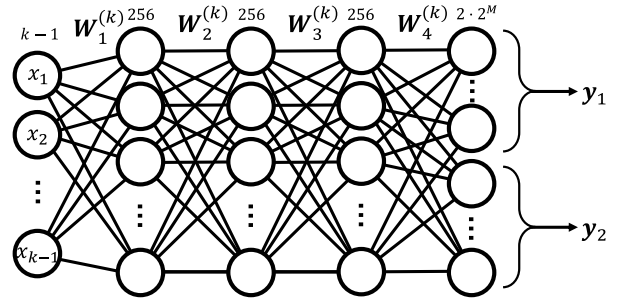}
    \caption{Fully connected neural network. }
    \label{fig:densenet}
\end{figure}

\end{itemize}


\section{Simulation results}

In this section we numerically examine the performance of the QFT sampler for several 
target distributions. 
For this purpose, the following criteria are used for the evaluation.

First, the cross entropy \eqref{cross entropy} is employed to see if the proposal distribution 
approaches toward the target. 
Note that in general the cross entropy does not decrease to zero even in the case when 
the proposal distribution coincides with the target. 
Hence, to evaluate the distance between the two probability distributions, we use the 
following Wasserstein distance: 
\[
      {\rm W}(P,Q) = \sup_{\lVert f \lVert_{L} \leq 1} 
             \Big\{  \mathbf{E}_{x \sim P} [f(x)] - \mathbf{E}_{x \sim Q} [f(x)]  \Big\},
\]
where the supremum is taken over $f$ contained in the set of functions satisfying 
the 1-Lipschitz constraint. 
Note that here the Kullback Leibler divergence is not used, because it is applicable 
only when the supports of the two distributions coincide.

We also use the average acceptance ratio to evaluate the efficiency of the sampler. 
Mathematically it is defined as the expectation value of Eq.~\eqref{acceptance ratio}, but 
in the simulation we simply take the ratio of the number of accepted samples to the 
number of all samples generated.


\subsection{1-dimensional case}

\begin{figure*}[htp]
    \centering
    \includegraphics[width=1.0\linewidth]{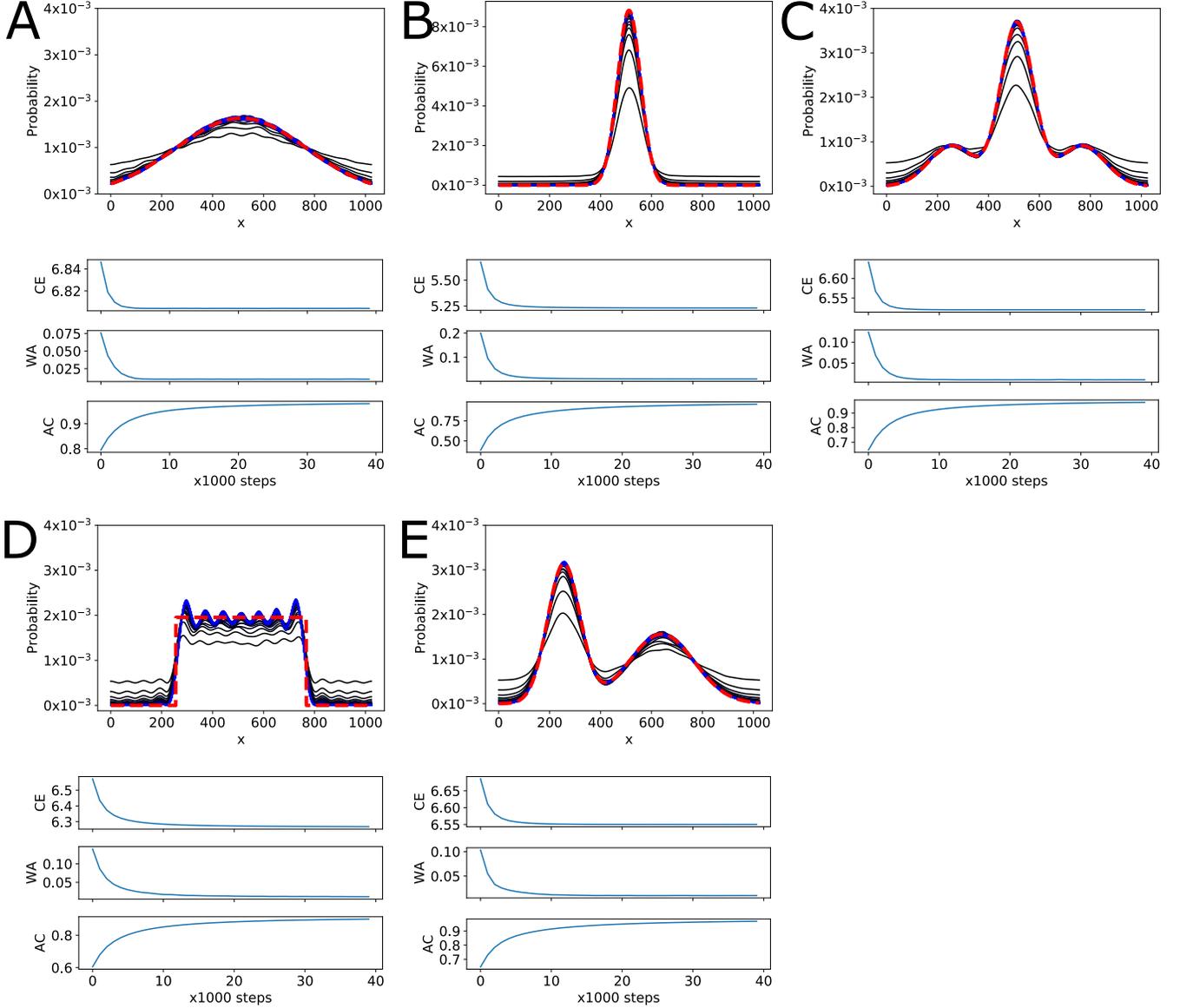}
    \caption{
    Performance of the 1-dimensional QFT samplers, for the five types of target distributions 
    shown with the red dotted line in the upper panel in each subfigures from A to G. 
    In the upper panel in each subfigure, some snapshots of the proposal distribution 
    generated from the QFT sampler are shown with solid lines. 
    In the bottom three panels in each subfigure, the convergence trend of the three types of 
    figure of merits are shown. 
    }
    \label{fig:r1d}
\end{figure*}

First we discuss the performance of the 1-dimensional QFT sampler. 
The QFT sampler is composed of $N=10$ qubits wherein $M=4$ qubits are used for 
parametrization; see Fig.~\ref{fig:m1}. 
The total learning step is 40000, where in each step $B=32$ samples are used to compute 
the gradient vector \eqref{compute gradient}. 
The learning coefficients in the momentum update rule \eqref{momentum method} are 
chosen as $\alpha=0.01$ and $\mu=0.9$. 
In the upper panels of Fig.~\ref{fig:r1d}, five types of target distributions (red broken lines) 
and snapshots of proposal distributions in each 1000 steps (black solid lines) as well as the 
convergent distributions (blue solid lines) are plotted. 
In the lower three panels in each subfigures, CE (Cross entropy), WA (Wasserstein distance), 
and AC (Average acceptance ratio) are plotted, demonstrating that in each case the proposal 
distribution $q_{\rm QFT}(x;\bm\theta)$ is approaching to the target $p(x)$, and accordingly 
AC increases. 
Note that in the case D, a wave-like structure still remains in the convergent proposal 
distribution, because of the absence of qubits corresponding to the high-frequency 
components. 
However, as mentioned in Sec.~\ref{intro}, this is not a serious issue in the framework 
of MH, which does not require a precise approximation of the target distribution via 
the proposal one but only needs a proposal distribution realizing relatively high acceptance 
ratio. 
In fact the lower panel of D in Fig.~\ref{fig:r1d} demonstrates about 30$\%$ improvement 
in the acceptance ratio.


\subsection{2-dimensional case}

\begin{figure*}[htp]
    \centering
    \includegraphics[width=1.0\linewidth]{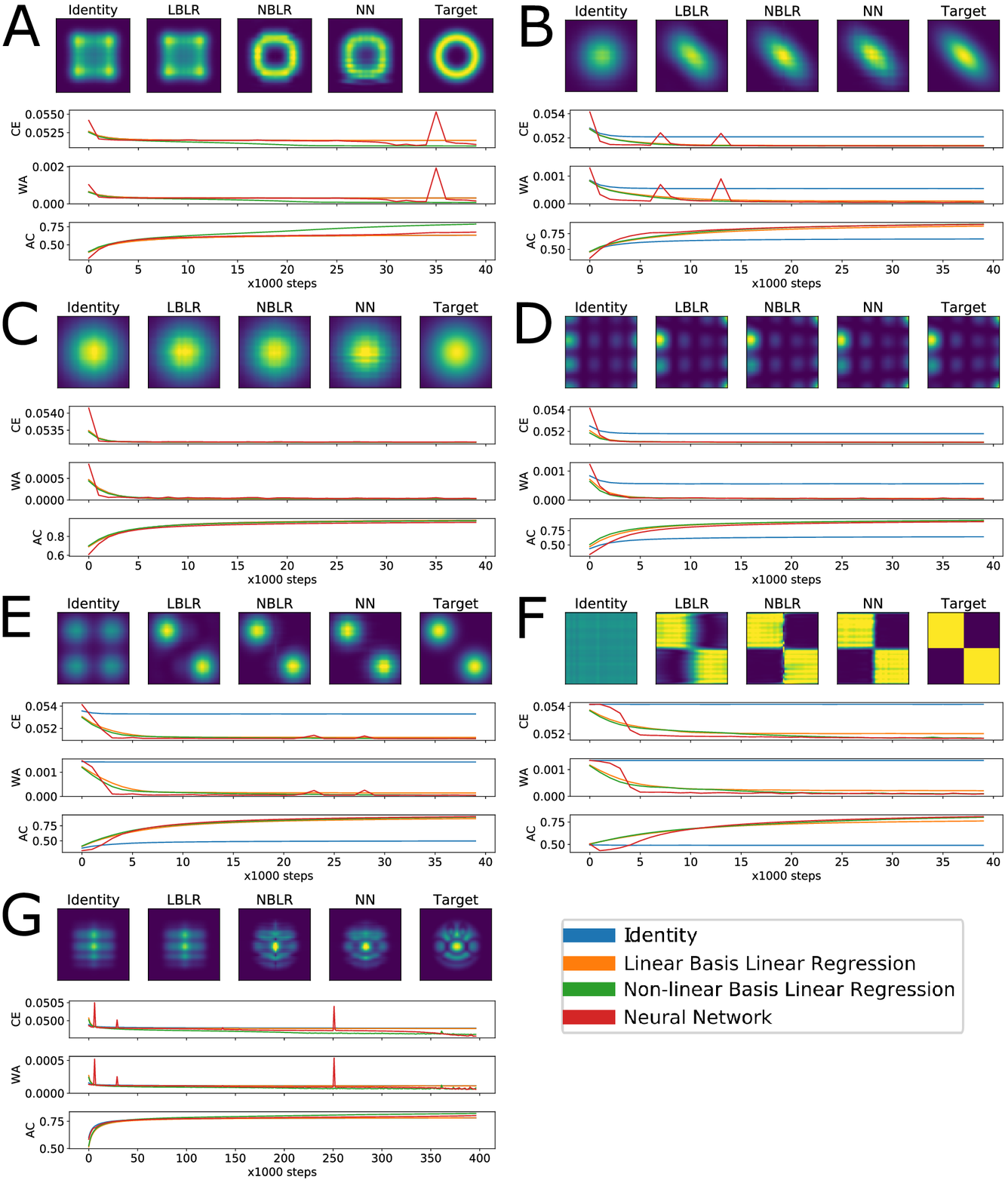}
    \caption{
    Performance of the 2-dimensional QFT samplers, for the five types of target distributions 
    shown in the upper right panel in each subfigures from A to G. 
    In the upper left four panels in each subfigure, the four proposal distributions generated 
    from the QFT sampler at the final step are shown. 
    In the bottom three panels in each subfigure, the convergence trend of the three types of 
    figure of merits are shown. 
    }
    \label{fig:r2d}
\end{figure*}

Next we study several 2-dimensional (i.e., $D=2$) target distributions, which are shown in 
the top right panels from A to G in Fig.~\ref{fig:r2d}. 
In this case, the proposal distribution \eqref{D dim proposal} is of the form 
\begin{equation}
\label{2 dim QFT sampler}
        q({\bm x};\bm\theta) 
          = q_{\rm QFT}(x_2, {\bm f}_2( x_1 ; \bm\theta_2 )) 
               q_{\rm QFT}(x_1, {\bm f}_1( \bm\theta_1 )), 
\end{equation}
where ${\bm f}_1( \bm\theta_1 )$ is set to be 
${\bm f}_1( \bm\theta_1 )={\rm Norm}(\bm\theta_1)$. 
As for ${\bm f}_2( x_1 ; \bm\theta_2 )$, we study the four models described in 
Sec.~\ref{sec:multi QFT sampler}; i.e., Id, LBLR, NBLR, and NN models. 
The learning coefficients of the momentum gradient method are set to 
$\alpha=0.01$ and $\mu=0.9$ for all the cases, except that $\alpha=0.001$ is chosen in 
the NN model for the cases from A to F. 
Also the two 1-dimensional QFT samplers in Eq.~\eqref{2 dim QFT sampler} are both 
composed of $N=10$ and $M=4$ qubits, for the cases from A to F, while $N=10$ and 
$M=5$ for the case G. 
The basis functions in the NBLR model ${\bm f}_2( x_1 ; \bm\theta_2 )$ are chosen as 
follows; 
\[
      \phi_1(x)=\bar{x}, ~~
      \phi_2(x)=\bar{x}^2, ~~
      \phi_3(x)=\bar{x}^3, ~~
      \phi_4(x)=\sqrt{|\bar{x}|}, 
\]
where $\bar{x}$ is defined as $\bar{x}=2x/(2^N-1)-1$. 
The total learning step is 40,000 for the cases from A to F and 400,000 for the case G. 
Finally, for all cases, $B=32$ samples are taken to compute the gradient in each step. 
With this setting, the proposal distributions at the final learning step corresponding to the 
four models (Id, LBLR, NBLR, and NN models from the left to right) are shown in the top 
panel of Fig.~\ref{fig:r2d}, and the change of the figure of merits (CE, WA, and AC) are also 
provided in the bottom panel.

First, as expected, the Id model can only approximate the distribution having no correlation in 
the space dimension, i.e., the case of C. 
On the other hand, the LBLR model acquires a distribution close to the target, for the cases 
B, D, E, F, in addition to C. 
The figure of merits, CE, WA, and AC, also reflect this fact. 
(The blue and orange lines in the cases A and C almost coincide.) 
It is notable that the simple LBLR model greatly improves the performance of the Id model.

To further handle the cases of A and G, some nonlinearities need to be introduced, as 
demonstrated by the NBLR and NN models. 
In particular, for all the cases from A to G, the NMLR model shows almost the same level of 
performance as the NN model, which is also supported by the figure of merits. 
Considering the fact that there are some jumps in WA and CE in the NN model, and the fact 
that the NN model costs a lot in the learning process, our conclusion is that the NMLR is 
the most efficient model in our case-study.


\subsection{Application to a molecular simulation}

\begin{figure*}[htp]
    \centering
    \includegraphics[width=1.0\linewidth]{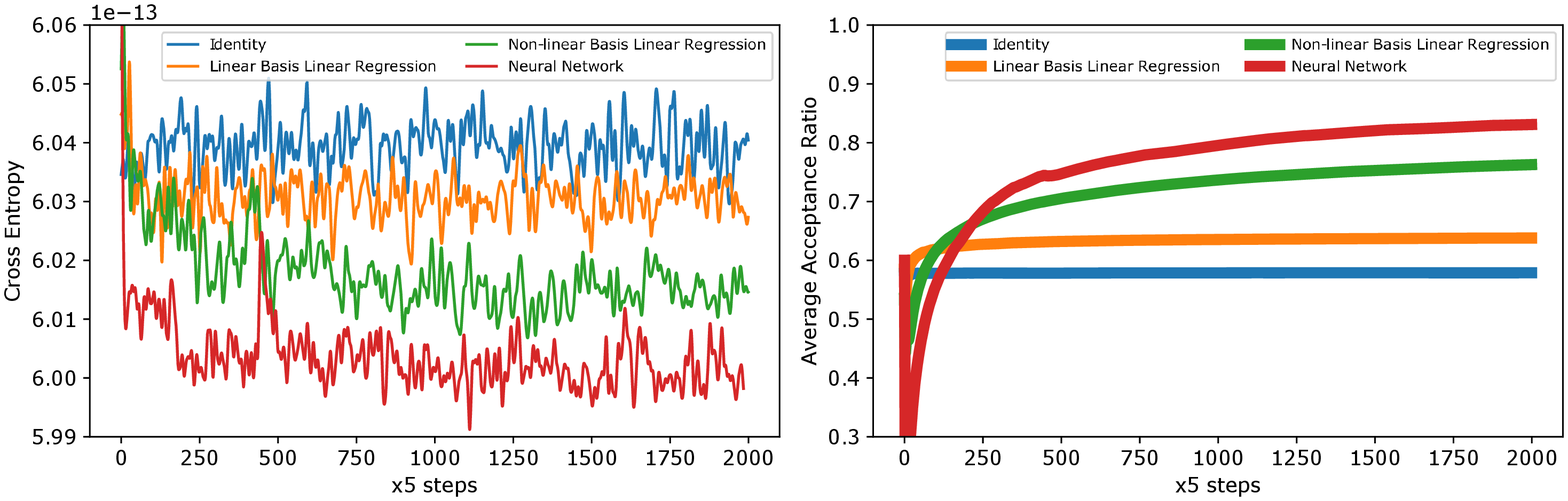}
    \caption{
    Convergence trend of the cross entropy \eqref{cross entropy} between the proposal 
    distribution generated from the QFT sampler to the target 6-dimensional Boltzmann 
    distribution (left) and the average acceptance ratio \eqref{acceptance ratio} (right). 
    }
    \label{fig:r6d}
\end{figure*}

The last case-study is focused on the stochastic dynamics of two atoms obeying the 
Lennard-Jones (LJ) potential field, which is often employed in the field of molecular 
simulation. 
This problem requires sampling from the Boltzmann distribution 
\[
     p(\bm r_1, \bm r_2) \propto {\rm exp} \Big\{ -\beta {\rm LJ} (\| \bm r_1 - \bm r_2 \| ) \Big\}, 
\]
where ${\rm LJ}(a) = a^{-12} - a^{-6}$ and $\beta=0.1$ is the inverse temperature. 
$\bm r_1 = [r_{1x}, r_{1y}, r_{1z}]^\top$ and $\bm r_2 = [r_{2x}, r_{2y}, r_{2z}]^\top$ are 
the vectors of position variables of each atom. 
We apply the 6-stages QFT sampler to generate a proposal distribution for approximating 
this target distribution.

The QFT sampler is configured by conditioning the samples in the order 
$r_{1x} \rightarrow r_{2x} \rightarrow r_{1y} \rightarrow r_{2y} \rightarrow r_{1z} \rightarrow r_{2z}$; 
for instance, $q_{\rm QFT}(r_{1y},  {\bm f}_{1y}( r_{1x}, r_{2x}; \bm\theta_{1y} ))$ is conditioned 
on $(r_{1x}, r_{2x})$. 
The output of the QFT sampler, $r_i \in \{0,1, \cdots ,2^N - 1\}$, is rescaled to 
$r_i/0.7(2^N - 1)$. 
We again employ the four models (Id, LBLR, NBLR, and NN models) as the conditioning functions, 
where ${\bm f}_1( \bm\theta_1 )$ is set to be 
${\bm f}_1( \bm\theta_1 )={\rm Norm}(\bm\theta_1)$. 
Each QFT sampler is composed of $N=10$ and $M=4$ qubits, and the total learning step is 
10000, whereas $B=1024$ samples are used to compute the gradient in each step. 
The learning coefficient is $\alpha=10$ for the Id, LBLR, and NBLR models, while $\alpha=0.001$ 
for the NN case; in each case the momentum method with $\mu=0.9$ is employed to update 
the parameters. 
The basis functions $\{ \phi_j^{(k)}(x_1, x_2, ..., x_{k-1}) \}$ for constructing the NBLR model 
are set to all the coefficients (except for the constant) of the third-order polynomial function 
$(1 + \bar{x}_1 + \bar{x}_2 + \cdots + \bar{x}_{k-1})^3$ with $\bar{x}= 2x/(2^N - 1) - 1$.

Figure \ref{fig:r6d} shows the change of CE and AC; WA was not calculated 
due to its heavy computational cost. 
We then find that, similar to the 2-dimensional case, the NBLR and NN models 
succeed in improving both CE and AC, although more learning steps are necessary 
compared to the previous cases. 
Note also that the Id and LBLR models are almost not updated, implying the validity to 
introduce nonlinearities in the model of QFT sampler.


\section{Conclusion}

This paper provided a new self-learning Metropolis-Hastings algorithm based on quantum 
computing and an important subclass of this sampler that uses the QFT. 
This QFT sampler is shown to be classically simulable, and the effectiveness of this quantum 
inspired method is supported by several numerical simulations. 
There are a lot of rooms to be investigated more, such as the choice of the optimizer 
and the conditioning function for constructing the multistage QFT sampler. 
In particular, in extending to the case of multi-dimensional distribution, a completely different 
schematic than the proposed multistage QFT sampler could be considered.

Although the QFT sampler offers a certain advantage over some classical sampling 
schemes as discussed at the end of Sec.~III A, of course, this quantum inspired algorithm 
is not what fully makes use of the true power of quantum computation. 
The real goal is to establish a genuine quantum sampler, which may provide a faster 
sampling with higher acceptance ratio than any conventional classical method. 
Such a direction is in fact found in the literature \cite{Temme,Moussa}, which are though 
far beyond the reach of current available devices. 
We hope that the present work might be a first step for bridging this gap.

\begin{acknowledgements}

This work was supported by IPA MITOU Target Program. 
KF and NY are supported by the MEXT Quantum Leap Flagship Program Grant Number 
JPMXS0118067394 and JPMXS0118067285, respectively. 

\end{acknowledgements}


\appendix


\section{Quantum circuit producing a multi-dimensional probability distribution}

The multi-dimensional probability distribution of random variables $\bm x=(x_1, \cdots, x_n)$ 
is simply obtained by measuring a state $\ket{\Psi}\in \otimes_{i=1}^n {\cal H}_i$ in the 
computational basis $\ket{\bm x}=\otimes_{i=1}^n \ket{x_i}$, i.e., 
$p(\bm x)=|\pro{\bm x}{\Psi}|^2$. 
The following is an example of 2-dimensional probability distribution such that the joint 
probability is explicitly represented via the conditional probability. 
Let $\ket{\phi_1}\otimes\ket{\phi_2}$ be an initial state on ${\cal H}_1\otimes {\cal H}_2$. 
We consider a quantum circuit of the form $\bm U=\sum_x \ket{x}\bra{x}\otimes U_x$, 
where $\ket{x}$ is a computational basis state of ${\cal H}_1$, and $U_x$ is a unitary 
operator conditioned on the value $x$; that is, $\bm U$ is a sum of controlled unitaries. 
Then the output probability distribution of the measurement result on the state 
$\ket{\Psi}=\bm U \ket{\phi_1}\otimes\ket{\phi_2}$, in the computational basis 
$\ket{\bm x}=\ket{x_1}\ket{x_2}$, is given by 
\[
     p(\bm x)
         = |\bra{x_1}\bra{x_2} \bm U \ket{\phi_1}\ket{\phi_2} |^2
                      = |\pro{x_1}{\phi_1}|^2 |\bra{x_2}U_{x_1}\ket{\phi_1}|^2. 
\]
Thus, in this case the joint probability $p(\bm x) = p(x_1, x_2)$ is explicitly given as a 
product of the conditional probability $p(x_2 \, | \, x_1)=|\bra{x_2}U_{x_1}\ket{\phi_1}|^2$ 
and $p(x_1)=|\pro{x_1}{\phi_1}|^2$.


\section{Classical simulation of the QFT sampler via adaptive measurement}

It is known that the a variety class of quantum circuits composed of the QFT can be 
simulated on a classical computer \cite{Browne}. 
This fact is indeed applied to our case, as described here. 
The point is twofold; one is that the input state to QFT is now of the form 
\[
        \ket{{\rm in}}= \ket{\psi(\bm \theta)} \otimes \ket{g^{N-M}}
           = [\theta_0, \ldots, \theta_{2^M-1}, 0, \ldots, 0]^\top, 
\]
where $\ket{\psi({\bm\theta})}= [\theta_0, \ldots, \theta_{2^M-1}]^\top$ is a relatively 
small quantum state whose entries can be efficiently determined. 
The other is that the circuit is terminated with the QFT part, meaning that the QFT is 
immediately followed by the measurement process.

\begin{figure}[b]
    \centering
    \includegraphics[width=1.0\linewidth]{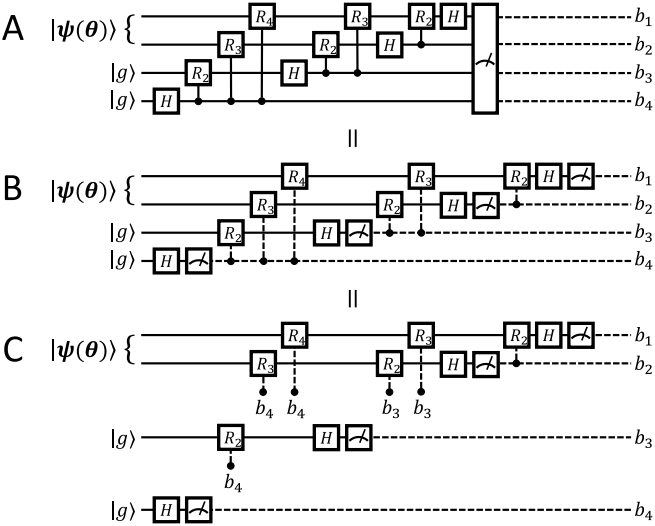}
    \caption{
    (A) Original circuit implementation of the QFT sampler on a quantum computer. 
    (B) Equivalent circuit implementation of the QFT sampler, in which all the controlled 
    rotation gates are replaced with the feedforward operations. 
    The solid and dashed lines denote quantum and classical operations, respectively. 
    (C) Equivalent circuit representation of the circuit B, emphasizing that the change of 
    the first 2-qubits (possibly entangled) state $\ket{\psi({\bm\theta})}$ and the bottom 
    two $\ket{g}$ states can be separately and adaptively tracked. 
    }
    \label{fig:ap1}
\end{figure}

Here we explain a detailed classical algorithm for simulating our circuit in the case of 
$N=4$ and $M=2$; 
hence $\ket{{\rm in}}= \ket{\psi({\bm\theta})} \otimes \ket{g} \otimes \ket{g}$ 
with $\ket{\psi({\bm\theta})}$ a two-qubits state. 
In this case, the original quantum circuit to implement the QFT sampler is shown in 
Fig.~\ref{fig:ap1}A. 
The main body of this circuit is QFT, which consists of the Hadamard gates denoted 
by $H$ and the controlled-$R_n$ gates which rotate the target qubit by acting 
$R_n = \exp{[-i2\pi\sigma_z 2^{-(n+1)}]}$ on it iff the control qubit is $\ket{e}=[0, 1]^\top$. 
The output of QFT is measured in the computational basis, producing the binary 
output sequence $b_1 b_2 b_3 b_4$ with $b_i\in\{0,1\}$. 
To classically simulate this circuit, we utilize the fact that, if the controlled rotation gate 
is immediately followed by a measurement on a control qubit, this process is replaced 
with the following feedforward type operation; 
that is, the control qubit is first measured, and, if the measurement result is ``1" the 
rotating operation acts on the target qubit. 
Repeating this replacement one by one from right to left in the circuit shown in 
Fig.~\ref{fig:ap1}A, we end up Fig.~\ref{fig:ap1}B. 
That is, all controlled-rotation gates are replaced with the 1-qubit rotation gate which 
depends on the antecedent measurement results. 

Using this classically controlled implementation, the QFT sampler can be simulated on 
a classical computer as described below. 
In the example considered here, the input 
$\ket{{\rm in}}$ is divided into at least three unentangled states 
$\{\ket{\psi({\bm\theta})}, \ket{g}, \ket{g}\}$. 
Hence, these three parts can be tracked separately and adaptively, as shown in 
Fig.~\ref{fig:ap1}C. 
In general, the change of states of the QFT sampler can be computed by repeatedly 
multiplicating $2^M$-dimensional matrices on $\ket{\psi({\bm\theta})}$ and 
$2$-dimensional matrices on $\ket{g}$ in the adaptive manner, leading that the 
total computational cost is of the order ${\mathcal O}(2^M + N)$. 
Therefore, the QFT sampler can be classically simulated as long as $2^M$ is a classically 
tractable number.


\end{document}